\documentclass[article,showpacs,preprintnumbers,amsmath,amssymb]{revtex4}
\usepackage{graphicx}
\usepackage{epstopdf}
\usepackage{rotating}
\usepackage{dcolumn}
\usepackage{bm}

\def\crossover{BEC-BCS crossover}
\def\li{$^6$Li }

\newcounter{romannumber}

\begin{document}
\draft

\title{Note on ``Collective Excitations of a Degenerate Gas at the BEC-BCS Crossover'', Phys. Rev. Lett. \bf{92}, 203201 (2004)}

\author{A. Altmeyer,$^{1}$ S. Riedl,$^{1,2}$ C. Kohstall,$^{1}$ M.J. Wright,$^{1}$ J. Hecker Denschlag,$^{1}$ and R. Grimm$^{1,2}$}
\address{$^{1}$Inst.\ of Experimental Physics and Center for Quantum Physics, Univ.\ Innsbruck,
6020 Innsbruck, Austria\\$^{2}$Inst.\ for Quantum Optics and Quantum
Information, Acad.\ of Sciences, 6020 Innsbruck, Austria}
\date{\today}
\begin{abstract}
We present a reinterpretation of our previous results on the radial
compression mode of a degenerate quantum gas in the BEC-BCS
crossover in \cite{Bartenstein2004b}. We show that our former data
are consistent with other experimental and theoretical work, when
the ellipticity of the optical trapping potential in
\cite{Bartenstein2004b} is properly taken into account .
\end{abstract}

\maketitle

\begin{figure}[!h]\center{
\includegraphics[width=16cm]{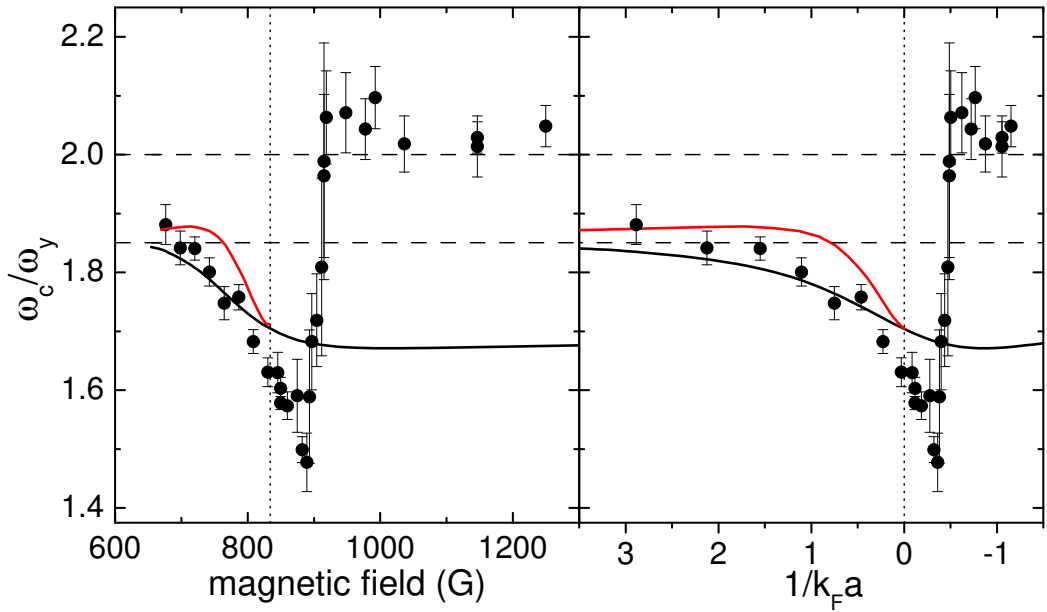}}
\caption{Normalized compression mode frequency
$\omega_{c}/\omega_{y}$ in the \crossover \,regime versus magnetic
field (left hand side) and interaction parameter $1/k_{F}a$ (right
hand side) \cite{Bartenstein2004b}. The lower theory curve is based
on a mean-field BCS model and the upper curve on a quantum
Monte-Carlo model. Both curves correspond to the theoretical data
presented in \cite{Astrakharchik2005}. The horizontal dashed lines
indicate the values for the BEC limit ($\omega_{c}/\omega_{y} =
1.851$ for $\zeta = 0.8$) and the collisionless limit
($\omega_{c}/\omega_{y} = 2)$. The vertical dotted line marks the
position of the Feshbach resonance at $834.1$G
\cite{Bartenstein2005}. \label{old_compression_bfield_kfa}}
\end{figure}

The radial compression mode of an optically trapped, ultracold \li
Fermi gas in the \crossover \,regime has been a focus of study of
experimental work performed at Innsbruck University
\cite{Bartenstein2004b,Altmeyer2006} and at Duke University
\cite{Kinast2004,Kinast2004a,Kinast2005}. In our most recent work
\cite{Altmeyer2006}, we reached a level of control  which allows us
to identify systematic effects in our measurements. With our new
knowledge of the system, we can reinterpret the previous data and
resolve the apparent discrepancy between \cite{Bartenstein2004b} and
\cite{Kinast2004,Kinast2004a,Altmeyer2006}.\\
The atoms are trapped by a single focused laser beam resulting in a
cigar-shaped trap geometry. In \cite{Bartenstein2004b}, we assumed
cylindrical symmetry along the $z$-axis of the trapping potential,
where the trap frequencies $\omega_{x}$ and $\omega_{y}$ in $x$- and
$y$-direction are equal. With this assumption we used $\omega_{y}$
as the relevant radial trap frequency $\omega_{r}$.\\
The experimental setup of \cite{Bartenstein2004b} was only capable
to resolve oscillations in $y$- and $z$-direction. With a new
imaging system along the $z$-axis we now get full access to the $x$-
and $y$-directions and are able to determine the two transverse
trapping frequencies individually. In contrast to the assumption of
cylindrical symmetry in \cite{Bartenstein2004b}, we found
significant ellipticity of the trap, being characterized by an
aspect ratio $\zeta = \omega_{x}/\omega_{y}$. For the experimental
trap setup of \cite{Bartenstein2004b} we found
an aspect ratio of $\zeta \approx 0.8$ \cite{newaspect}.\\
To calculate the frequency $\omega_{c}$ of the compression mode in
the elliptic trap, we start from the triaxial eigenfrequency
equation (e.g. \cite{Cozzini2003}) and neglect the weak confinement
in $z$-direction. This gives the collective mode frequencies
$\omega$ (compression mode and surface mode)
\begin{equation}
\label{cigar_shaped_coll_formula} \omega^{4} - (2 +
\Gamma)(\omega_{x}^{2} + \omega_{y}^{2})\omega^{2} + 4(\Gamma +
1)\omega_{x}^{2}\omega_{y}^{2} = 0,
\end{equation}
where $\Gamma$ is the polytropic interaction index. From equation
\eqref{cigar_shaped_coll_formula} the frequency of the radial
compression mode $\omega_{c}$ can be calculated \cite{Cozzini2003}.
This results in
\begin{equation}
\label{comp_mode_elliptic_cigar}
\left(\frac{\omega_{c}}{\omega_{y}}\right)^{2} = \frac{1}{2}\left(2
+ \Gamma\right)\left(1 + \zeta^2\right)+
\sqrt{\left(\frac{1}{2}\left(2 + \Gamma \right)\left(1 +
\zeta^{2}\right)\right)^{2} - 4\left(\Gamma + 1\right)\zeta^{2}},
\end{equation}
where $\omega_{c}$ is normalized to $\omega_{y}$, corresponding to
the way we presented our data in \cite{Bartenstein2004b}.\\
In Fig.~\ref{old_compression_bfield_kfa} the experimental data of
\cite{Bartenstein2004b} and theoretical data
\cite{Astrakharchik2005} corresponding to a mean-field BCS model
(lower curve) and a quantum Monte-Carlo model (upper curve), both
models assuming $\zeta = 0.8$, are shown. The same data set is
plotted versus the magnetic field (left-hand side) and the
interaction parameter $1/k_{F}a$ (right-hand side), where $a$
represents the atom-atom scattering length and
$k_{F}$ is the Fermi wave number. \\
In the BEC limit ($\Gamma = 2$) the data fit well with the
theoretically expected value of $1.85$. In the unitarity regime at
resonance ($\Gamma = \frac{2}{3}$), the experimental data also fit
well if one includes a small anharmonicity shift, which corrects
$\omega_{c}/\omega_{y} = 1.62(2)$ to $\omega_{c}/\omega_{y} =
1.67(3)$ \cite{Bartenstein2004b}. In the strongly interacting BEC
regime the data can be compared with the theoretical models using
equation \eqref{comp_mode_elliptic_cigar}. Above resonance, we see a
larger downshift in frequency until a jump to $\omega_{c}/\omega_{y}
\approx 2$ happens and the frequency remains constant.\\
In the strongly interacting BEC regime, at magnetic fields just
below the Feshbach resonance or $2 > 1/k_{F}a > 0.5$, the
experimental data points lie between both theoretical curves. In our
latest precision measurements \cite{Altmeyer2006}, the data clearly
support the quantum Monte-Carlo model and they also show a downshift
in frequency for increased temperatures. This is consistent with the
data presented in Fig.~\ref{old_compression_bfield_kfa}, taking into
account the relatively high temperature of the sample. Note that the
temperatures in \cite{Bartenstein2004b} are higher than in
\cite{Altmeyer2006} as the evaporation ramp was not optimized to
achieve deepest
temperatures and the timing sequence was not optimized to minimize heating.\\
At magnetic fields above the Feshbach resonance ($1/k_{F}a \lesssim
0$), the data show a significant downshift compared to the
theoretically expected values, which we cannot explain by the
elliptical trap. Also other experiments show a similar trend
\cite{Kinast2004a,dr_alex}. The proximity of the energy
corresponding to the collective mode frequency to the pairing gap
\cite{Combescot2004a} and thermal
effects may be possible explanations for this downshift.\\
At a magnetic field of about $900$G ($1/k_{F}a \approx -0.5$), the
normalized frequency shows a pronounced jump up to the value of
approximately $2$, which is expected in a collisionless Fermi gas.
For the collisionless oscillation along the $y$-axis, the
ellipticity of the trap is irrelevant. So the visibility of the jump
is clearly enhanced in an elliptic trap, as the lower frequency side
of the jump in an elliptic trap is downshifted compared to a
cylindrically symmetric trap. This jump marks the transition from
hydrodynamic to collisionless behavior.
\\
In conclusion, by taking into account the ellipticity of the
trapping potential, the results of \cite{Bartenstein2004b} now
essentially agree with other experimental results
\cite{Kinast2004,Kinast2004a,Altmeyer2006} and theoretical
predictions \cite{Astrakharchik2005}.\\
We thank Markus Bartenstein, Selim Jochim and Cheng Chin for their
contribution to the original results we reinterpreted here. We
acknowledge support by the Austrian Science Fund (FWF) within SFB 15
(project part 21). S.R.\ is supported within the Doktorandenprogramm
of the Austrian Academy of Sciences.


\begin{thebibliography}{39}
\bibitem[{newaspect()}]{newaspect}
\bibinfo{note}{In the new setup we use an aspect ratio of $\zeta = 0.94$.}
\end{thebibliography}
\end{document}